\documentclass[journal]{IEEEtran}
\usepackage{amsmath,graphicx,cite,epstopdf,amssymb,color}
\usepackage[numbers,sort&compress]{natbib}
\usepackage{subcaption}
\usepackage{multirow,graphicx,mathtools,adjustbox}
\usepackage{algpseudocode}
\usepackage[ruled, linesnumbered]{algorithm2e}

\begin{document}

\title{Integrated Energy Management for Operational Cost Optimization in Community Microgrids}

\author{Moslem~Uddin,
        Huadong Mo, and
         Daoyi Dong

	\thanks{Moslem Uddin is with School of Engineering Technology, The University of New South Wales, Canberra, ACT 2610, Australia (email: moslem.uddin.bd@gmail.com).}
\thanks{Huadong Mo is with School of Systems and Computing, The University of New South Wales, Canberra, ACT 2610, Australia (email: huadong.mo@unsw.edu.au).}
\thanks{Daoyi Dong is with Australian AI Institute, FEIT, University of Technology Sydney, Sydney, NSW 2007, Australia (email: Daoyi.Dong@uts.edu.au).}
}




\maketitle

\begin{abstract}
	
This study presents an integrated energy management strategy for cost optimization in multi-energy community microgrids (MGs). The proposed approach combines storage-based peak shaving, economic dispatch of diesel generators, and efficient utilization of renewable energy sources to enhance energy management in community MGs. The efficacy of the energy management system (EMS) was validated through a simulation case study for a rural Australian community.
The results demonstrate that the proposed EMS effectively reduces the peak energy demand by up to 43\%, lowers operational costs by 84.63\% (from \$189,939/year to \$29,188/year), and achieves a renewable energy utilization of 92.3\%, up from 47.8\% in the base system. Furthermore, the levelized cost of energy was reduced by 14.21\% to \$0.163/kWh. The strategy ensures an uninterrupted power supply during grid outages by utilizing DGs and battery energy storage systems. The environmental benefits included a 196.4\% reduction in CO$_2 $ emissions and 100\% reductions in CO, unburned hydrocarbons, and particulate matter. These findings validate the feasibility of the proposed EMS in achieving cost-effective, reliable, and sustainable energy management in community MGs. These findings contribute to the field by introducing a novel approach and demonstrating the practical feasibility of multi-energy MGs.  
\end{abstract}


\IEEEpeerreviewmaketitle

\section{Introduction}
%
%
%
%
\IEEEPARstart{T}{he} growing demand for sustainable energy solutions and the integration of renewable energy sources (RESs) have led to the growth of MGs as an efficient and resilient alternative to modern power grids. Among the different types of MGs, multi-energy community MGs have gained significant attention because of their ability to accommodate diverse energy resources and cater to the energy needs of the entire community \cite{houben2023optimal,roy2023comparison,ding2021economic}. 
This has the potential to improve power quality, increase energy reliability for critical loads, and enhance overall system efficiency  
\cite{uddin2023techno,gui2018passivity}. 
However, the unpredictable output of RESs and the complexity of meeting time-varying demands pose significant challenges \cite{gao2015energy,xu2022online}. Energy storage is a crucial technology for balancing demand and supply, and is widely used in MGs. Nevertheless, the coordination of storage systems, distributed RESs, and variable power demands remains a challenge \cite{he2019small}. 
Consequently, energy management (EM) for MG has attracted significant attention from both academics and industrial communities worldwide \cite{hao2019decentralized}.
Optimal management of an MG requires a systematic approach that balances energy supply and demand while maintaining reliability and reducing operating costs.

Recently, microgrid energy management (MEM) has been drawing an increasing interest
and remarkable progress has been made to develop more
advanced strategies. A stochastic framework-based MEM
was developed in \cite{gong2020secured}, where proposed EMS was investigated and analyzed in the context of a hybrid AC/DC MG.
However, techno-economic benefits are not investigated in
that study. 
Demand-response-based EMS for MG was presented in \cite{shen2016microgrid,yuan2022data}, but the effectiveness of demand response techniques depends on customer willingness. Robust techniques have been considered for MG EM applications, but they require significant computational time to design uncertainty bounds and solve problems 
\cite{bersani2016distributed,zhang2018robust}.  
A battery storage-based EMS was presented in \cite{thirugnanam2018energy}, but only solar photovoltaic (PV) was considered among the RESs. The incorporation of more RESs can help minimize the cost per kWh of energy production. A novel system operation technique for a small-scale direct current (DC)  MG integrated with RESs was demonstrated in \cite{liu2014system}, however, additional investigation is required to verify the effectiveness of the proposed method for AC MG. Scenario-based EMS have also been proposed 
\cite{nikmehr2017probabilistic,guo2016islanding,hadi2024real};
however, they are limited by the computational effort required to derive solution approaches for different probable scenarios.
Li et al. \cite{li2024distributed} proposed an optimal EMS that maximizes economic benefits while constraining system carbon emissions in islanded MGs. In \cite{manojkumar2022rule},  rule-based peak shaving (PS) scheme was introduced for an isolated MG to efficiently manage the peak load demand and ensure optimal operation of diesel generator (DG). However, it is essential to investigate their viability and potential issues in a grid-connected context. In addition, the combination of PS and economic dispatch has the potential to provide reliable and cost-effective power to communities. However, there is limited evidence to support this assertion.
Despite the limitations, these investigations provide valuable insights into EMS. Several studies may not necessarily be directly applicable to the multi-energy community MG context. Yet, they still hold significant value in offering lessons and guidance for the development and implementation of EMS in such settings. 
However, a more analytical and comprehensive approach is required to fully comprehend the potential of EMS in diverse energy scenarios. This involves exploring integral strategies, evaluating their performance in grid-connected environments, and incorporating price-based considerations into EM processes. Addressing these aspects will lead to a more comprehensive understanding of EMS and its potential for a cost-effective and reliable power supply to communities.

Therefore, this study aims to develop an EM framework for optimizing the operation of multi-energy MGs at the community level. To demonstrate the feasibility and effectiveness of this approach, a case study was conducted within a representative multi-energy community MG. In addition, this study discusses the implications of these findings for a broader field of multi-energy MG research and their potential to promote sustainable energy solutions. In summary, the significance and novelty of this study are as follows:

\begin{itemize}
	\item An integrated EMS is proposed to optimize the use of MG resources. 
	\item  A storage-based PS technique is introduced to mitigate peak demand complications in community MGs.
	\item Economic dispatch of the DG is implemented to provide uninterrupted power supply to the community.   
	\item The proposed EMS has been tested for an Australian rural community setting to verify its applicability.
	
\end{itemize}
The remainder of this paper is organized as follows. 
The methodology employed in this investigation is comprehensively described in Section II, the results obtained are presented, analyzed, and the findings are briefly discussed in Section III. Finally, Section IV presents the findings and conclusions drawn from the
investigation.

\section{Methods}\label{Sec II}

This study advances the initial investigation of storage-based EM approaches for multi-energy community MGs was  presented in \cite{uddin2024storage}. 
The earlier study introduced a preliminary MATLAB-based simulation framework for storage optimization and economic dispatch.  In contrast, this extended investigation introduced significant methodological enhancements. 
Initially, the HOMER Pro software is employed to design and optimize the system configuration by analyzing various scenarios and determining the most cost-effective and reliable combination of energy resources. This provides a foundational system design that considered different load profiles, resource capacities, and operational constraints. Building on this, a more comprehensive model is subsequently developed using MATLAB to explore EM issues, allowing for the incorporation of additional loads or profiles for further analysis. The MATLAB model facilitates a more in-depth investigation into the coordination of generator dispatch and peak load shaving, refining the operational strategy of the system for enhanced performance. 
The focus of this study is then directed towards validating the proposed approach through simulation case studies.

\begin{figure}[!tp]
	\centering
	\includegraphics[width= 3.5 in]{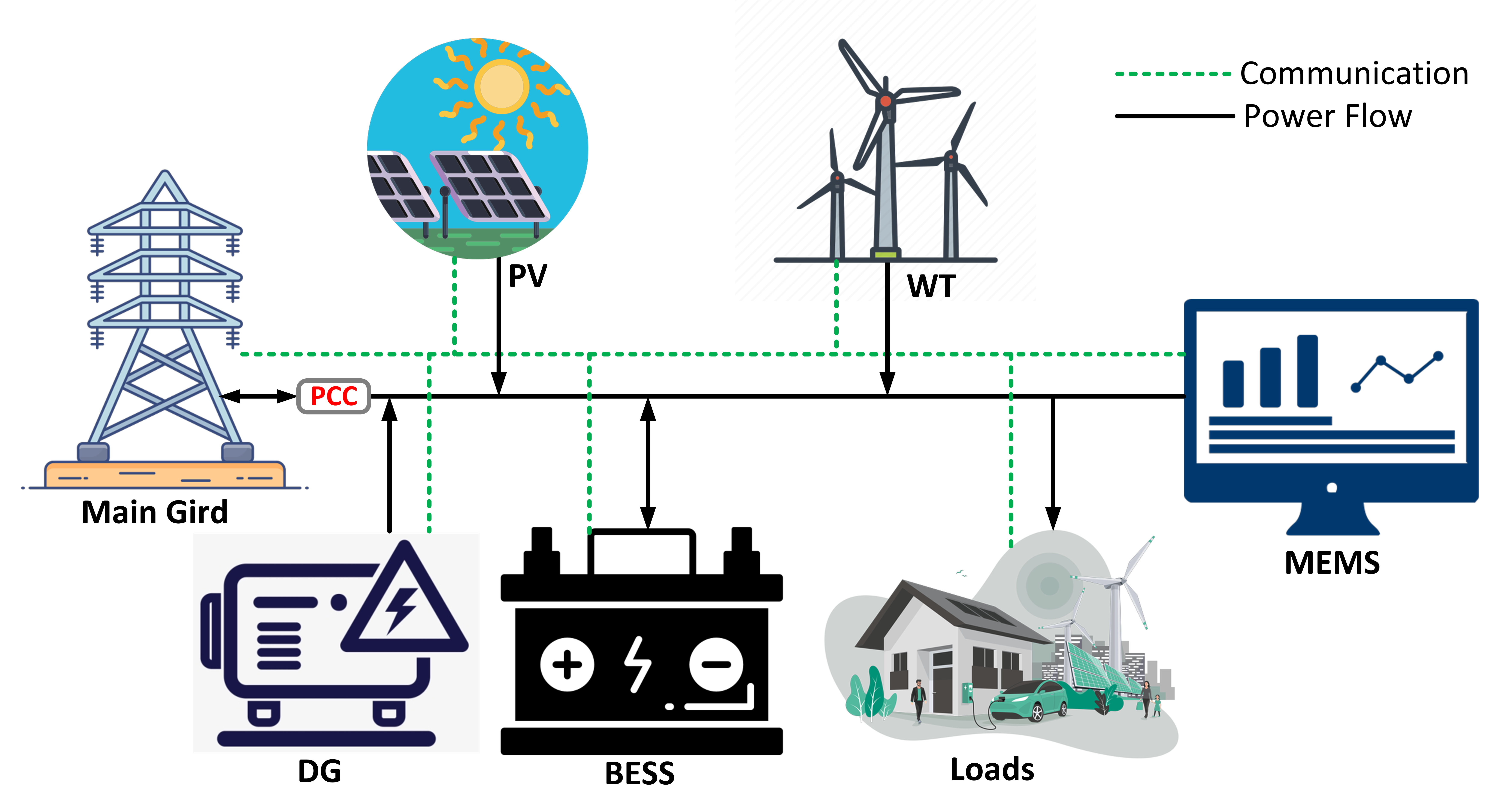}
	\caption{A Model of the Proposed MG Framework.}
	\label{fig_mg}
\end{figure}

\begin{algorithm}
	\SetAlgoCaptionLayout{centerline}
\caption{EMS-PS: PS Operation of EMS}
	\label{alg:peak_shaving}
	\SetAlgoLined
	\SetKwInOut{Input}{Input}
	\SetKwInOut{Output}{Output}
	
	\Input{Hourly load demand $P_d(t)$, real-time electricity price $P_e(t)$, Optimal size of PV $S_{pv}$, Wind $S_{wt}$, BESS $S_{bess}$, Grid Capacity $S_{grid}$, and DG $S_{dg}$.}
	\Output{Mitigate peak load complexity.}
	
	\For{each time step $t$}{
		\If{load exceeds a predefined threshold ($P_e^t$)}{
			Discharge the battery to reduce peak load and grid power imports\;
		}
	}
\end{algorithm}

\begin{algorithm}
	\SetAlgoCaptionLayout{centerline}
	\caption{EMS-GI: Grid interaction operation of EMS}
	\label{alg:grid_operation}
	\SetAlgoLined
	\SetKwInOut{Input}{Input}
	\SetKwInOut{Output}{Output}
	
	\For{each time step $t$}{
		\If{grid is available}{
			\If{$P_e(t) < P_e^t(t)$}{
				\If{$E_{sur} > 0$}{
					Charge the battery with $E_{sur}$\;
				}
				\Else{
					Discharge the battery to meet the load\;
				}
			}
			\Else{
				Discharge the battery to meet the load\;
				\If{$E_{sur} < 0$}{
					Import power from the grid (up to $GC_{max}$)\;
				}
				Calculate and record grid power import cost\;
			}
		}
		\Else{
			Grid outage\;
			\If{$E_{sur} > 0$}{
				Charge the battery with $E_{sur}$\;
			}
			\Else{
				Use the DG to meet the load (up to $P_{dg,max}$)\;
			}
			Calculate and record DG operating cost\;
		}
		
		Check SOC limits\;
		Calculate battery SOC\;
		Record battery energy level\;
	}
	
	Calculate cumulative imported energy, exported energy, and DG energy\;
	Calculate grid outage count and duration\;
	Calculate MG uptime\;
\end{algorithm}

\subsection{Test MG Design}
An MG typically comprises various power-generating resources, electrical loads, and centralized control units. In this study, the MG under investigation includes dispatchable DGs, non-dispatchable wind and solar PV systems, battery energy storage systems (BESS), and electrical loads.
The power is generated from renewable sources, such as solar and wind, depends on environmental conditions; therefore, their contributions to the MG cannot be accurately predicted or controlled.
The MATLAB model of the test MG comprises the dispatch control logic, BESS, and its control logic. A simplified structure of the proposed MG model is illustrated in Fig. \ref{fig_mg}. The system is designed to perform two functions: economic dispatch and load shaving, which require the implementation of two control logics.
As the test MG incorporates PV, wind turbine (WT), BESS, DG, and grid connection (GC), the power balance equation can be expressed as follows:
\begin{equation}\label{Eq-PowerBalance}
	P_d\left(t\right)+P_{pv}(t)+P_{wt}(t)+P_{dg}(t)\pm P_{gc}\left(t\right) \pm P_{bess}\left(t\right)=0
\end{equation}
where $ P_d\left(t\right) $ refers to MG power demand. The power generation from PV, WT and DG are denoted by $ P_{pv}(t) $, $ P_{wt}(t) $, and $ P_{dg}(t) $ respectively. $ P_{gc} $ refers to the power exported or imported from the grid. The charging and discharging powers of the BESS are represented by $ P_{bess} $.

\subsection{Proposed EMS}
Under the proposed EM approach, the MG operates in two modes: PS and grid-interaction operations. The peak-shaving application aims to decrease the highest electricity consumption by managing the battery charging and discharging cycles based on the electricity cost. In contrast, grid interaction operations involve managing the exchange of electricity between the MG and the GC, considering potential outages and price thresholds. These operations have the potential to lead to significant cost savings, enhance grid reliability, and promote efficient utilization of RESs. 

\subsection{PS Operation}
In this study, a storage-based peak-shaving strategy was implemented to reduce the peak electricity demand from the grid, which can result in decreased electricity costs and potentially prevent grid infrastructure upgrades. The details of the implemented PS strategy are presented in Algorithm \ref{alg:peak_shaving}. The implementation of PS in this study included the use of the following components:
\begin{itemize}
	\item \textit{PS Threshold}: PS commences with the establishment of a threshold value, denoted by $P_e^{t}$, expressed in dollars (\$). This threshold represents the uppermost price point at which the MG is prepared to purchase electricity without actively decreasing its consumption. This threshold is determined through real-time pricing information and a predetermined cost-saving strategy.
	\item \textit{Grid Price-Based Decision}: 
	The purpose of PS is to make decisions regarding the charging and discharging of the battery, based on the current electricity price at each time step. When the price of electricity is below a predetermined threshold, the MG utilizes excess renewable energy to charge the battery, which is an economical approach for storing energy for later use. Conversely, when the electricity price exceeds the threshold, the MG discharges the battery to meet the load, thereby reducing the power drawn from the grid during costly periods and resulting in cost savings. The charge/discharge decisions are determined using  Eq. \ref{Eq-PeakShaving}. 
	\begin{equation} \label{Eq-PeakShaving}
		P_e(t) = \begin{cases*}
			> P_e^t(t), & Discharge BESS,\\
			\leq P_e^t(t),              & Charge BESS. \\
		\end{cases*}
	\end{equation}
	
	\begin{table}[tp!]
		\centering
		\caption{Hourly electrical demand and electricity price (\$) for the case study community.}
		\label{tab_load&price}
		\begin{tabular}{cccl}
			\hline
			Time    & Load  & Electricity cost&\\ 
			(h)& (kW) & (Cants/kWh) &\\ \hline
			1 & 99.96 & 12.168& \\  		 
			2 & 73.5 & 12.24& \\
			3 & 70.56 &  8.496&	\\
			4 & 69.972  & 9.36 &\\
			5 & 92.904 &  11.52&	\\
			6 & 93.492  & 15.12 &\\
			7 & 102.9  & 22.896& \\
			8 &  161.7 & 22.752& \\
			9 & 135.24 & 25.92 &\multirow{5}{*}{\hspace{-3em}$\left.\begin{array}{l}  \\
					\\
					\\
					\\
					\\
				\end{array}\right\rbrace$ {Peak  price}}  \\
			10 & 88.2 & 25.2& \\
			11 & 87.024 &27 & 	\\
			12 & 87.024  &26.64 &  \\
			13 & 88.2  &26.424 &  \\
			14 & 170.52  &21.6&  \\
			15 & 161.7  &23.184& \\
			16 & 164.052  &22.968 &  \\
			17 & 182.28 &23.256&  \\
			18 & 185.22  &26.28 & \multirow{4}{*}{\hspace{-3em}$\left.\begin{array}{l}  \\
					\\
					\\
					\\
				\end{array}\right\rbrace $ {Peak  price}}  \\
			19 & 211.68	  &28.872	&  \\
			20 & 226.38  &26.496 &  \\
			21 &  235.2 &28.08&  \\
			22 & 129.36  &20.52 &  \\
			23 & 128.184 &18&  \\
			24 & 94.08  &12.6 &  \\		 
			\hline
		\end{tabular}
	\end{table}

	\item \textit{Battery State of Charge (SOC) Limits}: To maintain safe and efficient battery operation, SOC constraints are established. Constraints $ SOC_{min} $ and  $ SOC_{max} $  define the minimum and maximum allowable SOC levels, respectively.	If the SOC falls below the lower limit, excess renewable energy is utilized to charge the battery until it reaches the lower limit. If the SOC exceeds the upper limit, excess energy is not stored, and the battery remains at the upper limit. When the SOC of the battery is within the limits specified by the upper and lower thresholds, the battery control logic allows the battery to undergo charge or discharge operations. These operations are governed by Eq. \ref{Eq-SOC}.
	\begin{equation}\label{Eq-SOC}
		SOC = \begin{cases*}
			SOC_{max}, & Declaine charging,\\
			SOC_{min},              & Declaine discharging,\\
			otherwise,                    & Charge/Discharge.
		\end{cases*}
	\end{equation}
	
\end{itemize}

\subsection{Grid Interaction Operations}
Grid interaction operations pertain to the interaction between the MG and the external grid, which involves the import and export of electricity. Details of the implemented approach are presented in Algorithm \ref{alg:grid_operation}. The following outlines how grid interaction is managed under this operation.
\begin{itemize}
	\item \textit{Grid Power Import and Export}:  The MG can engage in the import or export of electricity to or from the grid, depending on the sufficiency of the renewable energy generation and battery capacity to meet the load. If there is a surplus of renewable energy, the MG exports excess energy to the grid, whereas if there is a deficiency, it imports electricity from the grid. The surplus energy is calculated using  Eq. \ref{Equation_Imp-Exp}.
	\begin{equation}\label{Equation_Imp-Exp}
	E_{\text{sur}} = \sum_{t=1}^{T} \left(P_{pwb}(t) - P_{\text{d}}(t) \right) \Delta t
	\end{equation}
	where, $ P_{pwb}(t) $ the aggregate power output from PV, WT, and BESS at time t.
	
	\item \textit{DG Dispatch}: 
	In the event of a grid outage, the MG operates autonomously, thereby offering a critical level of energy resilience. Its autonomous operation enables the MG to fully utilize its available resources, such as DGs, to their maximum designated capacity. By strategically managing its energy assets in this manner, the MG minimizes disruptions and ensures uninterrupted power supply during grid disturbances, ultimately enhancing grid reliability. In this case, the power balance equation (Eq. \ref{Eq-PowerBalance}) can be rewritten as follows:
	\begin{equation}
		P_d\left(t\right)+P_{pv}(t)+P_{wt}(t)+P_{dg}(t)\pm P_{bess}\left(t\right)=0.
	\end{equation}
	
	\item \textit{Electricity Price Threshold}: In this operational mode, the MG employs a price threshold akin to PS  to determine whether to charge or discharge the battery when it is connected to the grid. When the electricity price falls below this threshold, the MG charges its battery. Conversely, when the price exceeds the threshold, the battery is discharged to fulfill the demand.
\end{itemize}

\subsection{Case Study}
The proposed EMS was tested for the rural community of Central Tilba in New South Wales, which comprises 288 residents. Despite being recognized as one of Australia's most significant cultural locations, the region still experiences unreliable and costly access to electrical power.  The average daily power consumption was approximately 3139.3 kW/day, and the peak demand was approximately 235.2 kW. Table \ref{tab_load&price} shows the load profile of the rural community. The daily electrical loads also followed an oscillating distribution, with a higher power demand during the day and a lower power demand at night, with a peak load occurring between 5:00 pm and 9:00 pm. 
One of the features of an MG is its ability to purchase electricity and sell it back to the grid utility at a real-time price, denoted as $P_e(t)$ (\$/kWh). This price fluctuates in real time to efficiently manage power demand by shifting loads from peak hours to off-peak hours. The rate is determined by wholesale market prices, which change based on power demand, and the peak demand indicates high electricity utilization. The indicative real-time electricity rate is presented in Table \ref{tab_load&price}. The unit of the rate is expressed in USD per kilowatt-hour. The rates are set to correspond with the future Australian electricity market, although the variation is taken from the real-time prices.

\begin{table}[tp!]
	\centering
	\caption{Technical characteristics and economic data for MG components.}
	\label{tab:tec_eco_data}
	\begin{tabular}{llll}
			\hline
			Components & Characteristics   & Values & Unit  \\ \hline 
			\multirow{6}{*}{PV } &  Nominal power, $ P_{pv}$ &  1kW  & kW  \\  
			& Derating factor   & 80 & \% \\  
			& Capital cost   & 1300 & \$/kW\\ 
			& Replacement cost  & 1300 & \$/kW \\
			& O\&M cost    & 10 & \$/kW\\ 
			& Lifetime  &  20 & Year \\  	&&&\\

			\multirow{7}{*}{WT } &  Nominal capacity, $ P_r $ &   3&kW \\  
			& Cut-in wind speed,  $V_ {ci} $  & 4 & $ m/s^2 $ \\  
			& Cut-out wind speed, $ V_{co} $   & 24 & $ m/s^2 $\\
			& Hub height, h  & 15 & m \\
			& Capital cost    & 2300& \$/kW\\
			& O\&M cost  & 207 & \$/kW/year \\  
			& Lifetime & 20 & Year \\  	&&&\\
			
			\multirow{4}{*}{DG } & Nominal capacity, $ C_{deg} $   & 60 &kW  \\  
			& Capital cost   & 400 &\$/kW \\  
			& Replacement cost   & 400 & \$/kW \\  
			& O\&M cost   & 0.03& \$/h/kw\\  	&&&\\

			\multirow{8}{*}{BESS } & Nominal capacity, $ C_{batt} $  &  1&kWh  \\  
			& Nominal voltage, $ V_{batt}  $ &  24 & V \\  
			& Roundtrip efficiency, $\eta_{RT} $  &  90& \% \\ 
			& DoD &  80 &\% \\ 
			& Capital cost  &  700 &\$/kW \\ 
			& Replacement cost  & 700&\$/kW  \\
			& O\&M cost  &  10 & \$/year/kWh\\ 
			& Lifetime & 10 & Year \\   
			&&&\\
			
			\multirow{6}{*}{Converter } &  Nominal Capacity & 1 &kW \\  
			& Conversion efficiency, $ \eta_{inv} $  & 95 &\%  \\  
			& Capital cost  & 300 &\$/kW \\ 
			& Replacement cost   & 300 &\$/kW  \\ 
			& O\&M cost    &  0 & \$/year \\  
			& Lifetime   & 15 & Year\\  \hline
			
		\end{tabular}
\end{table}

\subsection{Simulation Setup}
The simulation of the MG's operation is conducted using MATLAB and HOMER Pro, two advanced tools widely recognized for their robust capabilities in modeling and optimizing energy systems. MATLAB is employed to perform detailed technical simulations encompassing the system dynamics, control strategies, and response analysis under various operational conditions. HOMER Pro provides a framework for economic analysis and optimization, facilitating the assessment of the system feasibility under diverse economic scenarios. These tools enable a comprehensive simulation of energy production, consumption, storage, and financial metrics, ensuring a thorough evaluation of the MG system.

Table \ref{tab:tec_eco_data} presents the fundamental technical characteristics and economic data for each component of the MG system, , as detailed in \cite{uddin2023techno}. This table encompasses a range of crucial parameters, including nominal capacities, operational costs, and lifetimes for photovoltaic modules, wind turbines, DGs, battery storage, and power converters. The data provided form the basis for the analysis of system performance, investment costs, and operational efficiencies, enabling a comprehensive assessment of the feasibility and sustainability of the MG.

\subsection{Scenario-Based Performance Evaluation}
In contrast to the preliminary analysis presented in \cite{uddin2024storage}, this study incorporates a comprehensive multi-scenario simulation framework to enhance the accuracy of the system modeling. Furthermore, the proposed EMS has been evaluated under a broad range of operational scenarios, including economic stress tests and extended grid outages, to assess its robustness and scalability.

\begin{itemize}
	\item \textbf{High Demand Scenario (S1):} Simulating the MG’s performance during extreme weather conditions that result in increased energy demand, such as severe cold periods or heat waves.
	\item \textbf{Low Renewable Output Scenario (S2):} Assessing system reliability when solar and wind power outputs are significantly reduced due to adverse weather conditions, such as storms, floods and bushfires.
	\item \textbf{Grid Failure Scenario (S3):} Evaluating the MG's capability to operate independently (islanding) during a grid outage, caused by a combination of technical and human-related issues. This scenario focuses on evaluating the capacity of the system to meet the energy requirements of the community solely through renewable sources, distributed generation, and stored energy.
	\item \textbf{Economic Stress Test (S4):} Analyzing the financial viability of the MG under fluctuating fuel prices.
\end{itemize}

These scenarios are specifically selected to elucidate the strengths and identify the potential limitations of the proposed EM system, thereby providing a comprehensive framework for evaluating its overall efficacy and resilience.

\section{Results and Discussion}
In this section, the simulation results are presented to demonstrate the performance of the proposed MEM strategy.

\begin{figure}[tp!]
	\centering
	\includegraphics[width= 3.7 in]{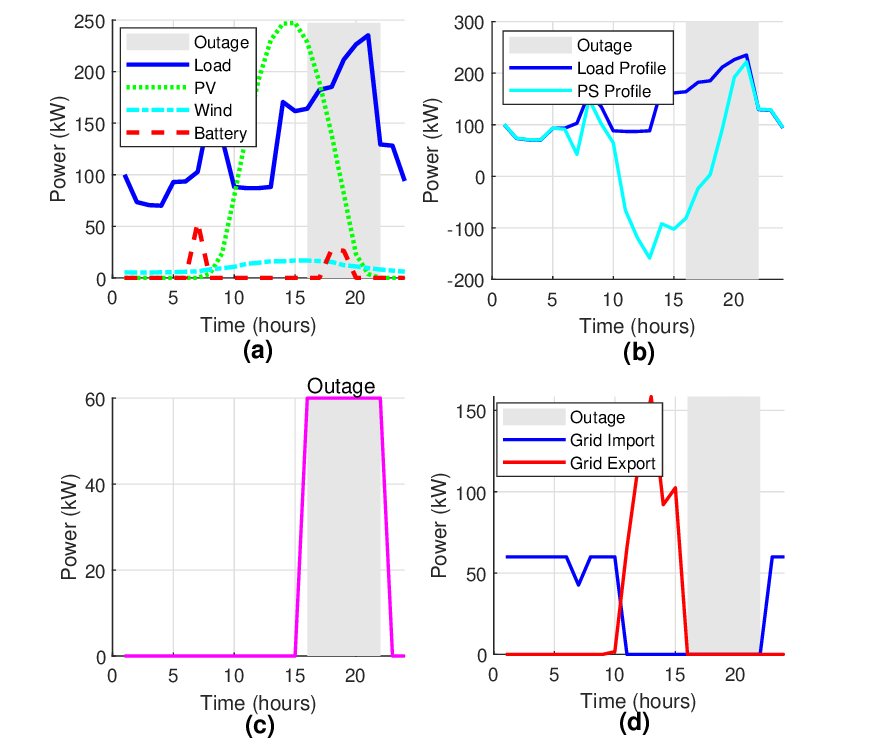}
	\caption{Comprehensive overview of MG energy dynamics over a 24-hour period: (a) Load profile in conjunction with PV, wind and BESS generation profiles, showing renewable integration. (b) Comparison of original and managed load profiles, demonstrating the impact of EM strategies, including battery storage operations. (c) DG output, illustrating its role in stabilizing the MG during periods of low renewable generation. (d) Grid interactions depicting both import and export activities, reflecting the MG's capacity to adapt to fluctuating energy demands and supply conditions.}
	\label{fig_r1}
\end{figure}

\subsection{Performance Evaluation}
The load profile of the studied multi-energy MG community is shown in Fig. \ref{fig_r1} (a). The load exhibited considerable variability with peak demand periods occurring intermittently throughout the day. The maximum load demand attained 235.2 kW, while the minimum demand decreased to 69.972 kW.  Fig. \ref{fig_r1} (a) also illustrates the renewable generation for the case study MG, exhibiting a comparable degree of variability throughout the day. Renewable generation achieved a maximum of 263.3 kW, whereas the minimum renewable generation decreased to 5.18 kW.
The proposed EM system demonstrated significant operational efficiency and reliability in a simulated multi-energy MG community. The simulation results depicted in Fig. \ref{fig_r1} (b) indicate substantial reductions in peak energy demand by up to 43\% during critical periods, primarily between 7:00 pm and 9:00 pm, when energy consumption typically reaches its maximum due to increased domestic activities. The proposed PS  capability not only reduces energy costs but also mitigates strain on the grid infrastructure.

The function of DG as a backup power source was critically evaluated, as illustrated in Fig. \ref{fig_r1} (c), demonstrating its reduced activation frequency due to enhanced battery management and renewable integration. During the simulated grid outages, the DG is activated solely in scenarios where a grid outage occurs, and the combined storage and renewable generation are insufficient to meet the demand, contributing to an overall reduction in fuel consumption compared with conventional systems lacking advanced energy storage solutions.
Moreover, when a critical timeline commenced  at 16:00, coinciding with an increase in the power demand, as depicted in Fig. \ref{fig_r1} (a), an unanticipated grid outage occurs. Although the BESS contributed a portion of the required power at this junction, it could not fully satisfy the escalating load. This partial contribution was attributed to the depleted energy reserves in the BESS, which had a battery energy level of zero, indicating its inability to provide complete support during the outage. Contingency DG was activated in response to this power supply deficit, as shown in Fig. \ref{fig_r1} (c). This generator was strategically employed to address the discrepancy between the prevailing demand and combined capacity of the RESs and BESS. It is essential to emphasize that the generator is utilized judiciously in accordance with its intended function as an emergency backup power source to ensure the uninterrupted provision of electricity under adverse circumstances.

\begin{figure}[!tp]
	\centering
	\includegraphics[width= 3.7 in]{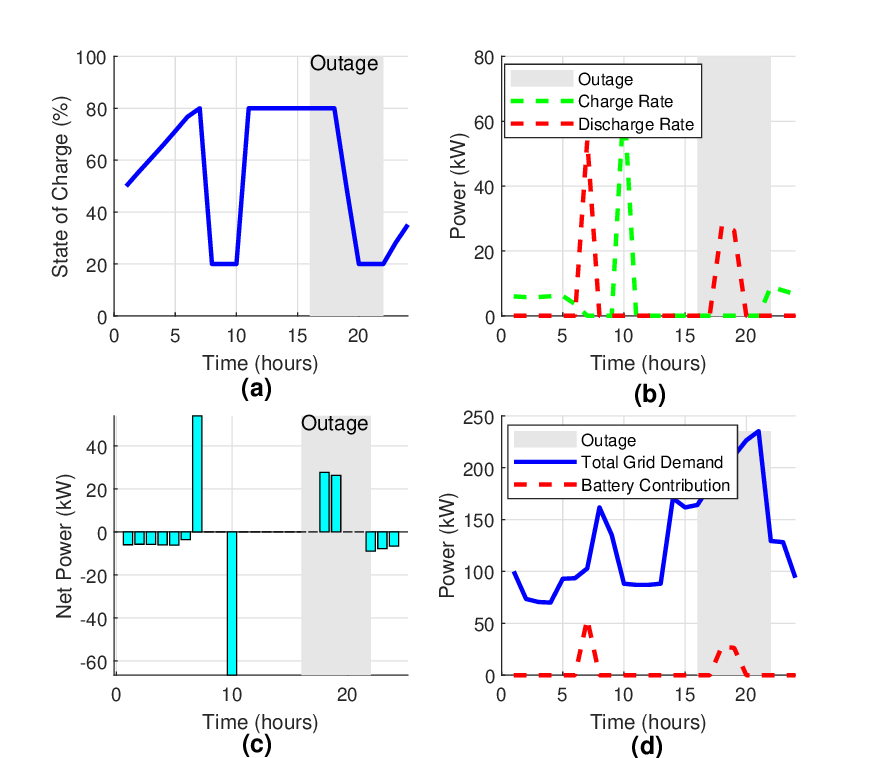}
	\caption{Analysis of battery and grid interactions over time. (a) Battery SOC depicting the percentage of charge throughout the simulation period. (b) Battery charging and discharging rates indicating power inflow and outflow from the battery. (c) Net battery power contribution illustrating the balance between charging and discharging. (d) Comparison of total grid demand and battery contribution, demonstrating the battery's role in meeting energy requirements.}
	\label{fig_r2}
\end{figure}

 Fig. \ref{fig_r1} (d) illustrates the interaction between the MG and external power network. The import curve delineates the instances when the MG draws electrical power from the grid to compensate for energy deficits. These imports typically occur during periods when the battery's SOC falls below 20\%, ensuring that a sufficient quantity of energy is available during intervals of low renewable energy generation or high demand. Conversely, the export curve depicts instances in which the MG exports excess renewable energy to the grid, functioning as a distributed energy resource.

\subsection{Energy Storage Utilization}
The BESS played a pivotal role in the EM. To minimize the load variability, the BESS was operated as illustrated in Fig. \ref{fig_r2}. The SOC levels of the BESS were optimized to be maintained between 20\% and 80\% throughout the operational cycle, as shown in Fig. \ref{fig_r2} (a). This optimization ensured a stable energy supply, while also enhanced the longevity of the storage units. 
During a simulated power grid failure, the MG demonstrated enhanced resilience against network instabilities and blackouts. The system successfully maintained its load independently for up to 6$ h $ without any significant decrease in efficiency, thereby illustrating its improved capacity to withstand grid fluctuations and outages.
 Fig. \ref{fig_r2} (b) illustrates the energy dynamics of the battery within the MG. This demonstrates the significance of implementing the proposed EM strategy, which effectively balances energy supply and demand.  Fig. \ref{fig_r2} (c) reveals that the net power (kW) of the BESS fluctuates as it interacts with renewable generation and load requirements. During periods of excess renewable energy generation, the battery stored the energy and reached a maximum capacity of 58 kW. Conversely, the battery discharges to satisfy the load when renewable generation is insufficient. Efficient management of the BESS is essential to ensure an uninterrupted power supply and minimize grid dependence.

The comparative analysis of the grid demand and battery contribution is presented in Fig. \ref{fig_r2} (d), which demonstrates a well-coordinated interaction between the battery storage and grid requirements. The power contribution of the battery exhibits an inverse correlation with grid demand patterns, illustrating its efficacy in addressing demand gaps when the grid power is insufficient or cost-prohibitive. This strategic utilization of battery resources not only enhances grid stability, but also facilitates a more sustainable and economically viable operation of the energy system by mitigating the necessity for expensive peak power generation from conventional sources.

\begin{table}[tp!]
	\centering
	\caption{Economic Analysis Results: Comparing Proposed and Base Systems}
	\begin{tabular}{|l|c|c|c|}
		\hline
		Metric              & Base System & Proposed System & \% Change \\ \hline
		NPC                 & \$4.47M     & \$1.69M         & -62.19\%  \\ \hline
		LCOE                & \$0.190/kWh & \$0.163/kWh     & -14.21\%  \\ \hline
		Operating Cost      & \$189,939/year & \$29,188/year  & -84.63\%  \\ \hline
		CAPEX               & \$1.18M        & \$1.18M        & 0.0\%     \\ \hline
		Renewable Fraction  & 47.8\%         & 92.3\%         & +93.10\%  \\ \hline
	\end{tabular}
	\label{tab:economic_analysis_results}
\end{table}

\subsection{Economic Benefits}

Table \ref{tab:economic_analysis_results} summarizes the quantitative economic benefits of implementing the integrated EM compared to the base system. 
It presents a comprehensive comparison of several critical financial metrics: Net Present Cost (NPC), Levelized Cost of Energy (LCOE), Operating Costs, Capital Expenditures (CAPEX), and Renewable Utilization (RU). Each metric demonstrates the substantial economic advantages of the proposed system, providing clear insights into its financial viability and potential cost savings achievable over the operational lifespan of the system.

The proposed MEM exhibited a significant reduction in NPC compared with the base system, indicating a decrease of 62.19\%. This substantial reduction is indicative of the long-term cost savings associated with the MG, despite its higher initial and operating costs. This decrease can be attributed to the enhanced efficiency of the MG and the integration of renewable energy sources, which likely offset the initial higher costs over time. The LCOE of the MG was notably lower than that of the base system, with a reduction of 14.21\%. This finding underscores the cost-effectiveness of MG in generating energy, making it an economically viable option for stakeholders. A lower LCOE reflects the efficient use of technology in the MG, which optimizes energy production costs per unit and contributes to overall sustainability and affordability.

The analysis revealed a significant 84.63\% decrease in operational expenses, with annual costs dropping from \$189,939  in the base system to \$29,188/year in the proposed system. This substantial decrease indicates improved operational efficiency and potentially the implementation of a more advanced strategy in the proposed system. This reduction not only enhances the economic viability of the system but also increases its appeal to potential investors and operators.

The CAPEX for both the proposed system and the base system were identical. This configuration was designed to isolate and investigate the benefits of the proposed EM system. The equivalence in initial setup costs is particularly significant, as it demonstrates that integrating a more sustainable and technologically advanced MG does not necessitate an additional initial financial outlay compared to a conventional system. For stakeholders, this implies that the adoption of the MG can be implemented without deterring higher upfront costs, which is a common barrier in the deployment of new technologies. This cost parity could significantly facilitate the adoption of MGs, especially in regions where financial constraints are a major obstacle to infrastructure advancement.

There is a substantial increase in the renewable fraction of the MG, up by 93.10\%, reflecting a strategic shift towards more sustainable energy sources. This shift contributes to reducing dependency on non-renewable energy while aligning with global trends and regulatory policies favoring renewable energy, thus enhancing the system's compliance with environmental standards. 

\begin{table}[tp]
	\centering
	\caption{Annual Emissions Reduction and Impact Analysis}
	\label{tab:emissions_impact}
	\begin{tabular}{|l|c|c|c|}
		\hline
		\textbf{Metric} & \textbf{Base System} & \textbf{Proposed System} & \textbf{Reduction (\%)} \\ \hline
		CO$_2$ (kg/yr) & 315,909 & -310,462 & 196.4\% \\ \hline
		CO (kg/yr) & 137 & 0 & 100\% \\ \hline
		UH (kg/yr) & 5.53 & 0 & 100\% \\ \hline
		PM (kg/yr) & 0.546 & 0 & 100\% \\ \hline
		SO$_2$ (kg/yr) & 1,332 & -1,346 & 201.1\% \\ \hline
		NO$_2$ (kg/yr) & 638 & -658 & 203.1\% \\ \hline		
		RU (\%) & 47.8 & 93.2 & 95\% \\ \hline
	\end{tabular}
\end{table}

\begin{table*}[tp!]
	\centering
	\caption{Scenario-Based Performance Evaluation Results of the MG}
	\begin{tabular}{|l|l|l|c|c|c|c|c|}
		\hline
		\multicolumn{2}{|c|}{\textbf{Scenario}} & \textbf{Description} & \textbf{Condition Changes} & \multicolumn{4}{c|}{\textbf{Impact}} \\ \cline{5-8}
		\multicolumn{2}{|c|}{} & & & \textit{Reliability} & \textit{Economic} & \textit{Efficiency} & \textit{Stakeholder}  \\ \hline
		\multicolumn{2}{|l|}{S1} & Increase in demand & +5\% & High & Moderate & High & Higher energy bills for consumers\\ \hline
		\multirow{2}{*}{S2} & PV & Reduction in PV output & -20\% & Moderate & High & Moderate & Increased reliance on grid\\ \cline{2-8}
		& Wind & Increase in wind output  & -40\% & Moderate & High & Moderate & Higher utility costs\\ \hline
		\multicolumn{2}{|l|}{S3} & Grid outage  & 6$ h $ Outage &High & Low & High & Increased reliance on DG \\ \hline
		\multicolumn{2}{|l|}{S4} & Fuel price fluctuation & +100\% Fluct. & Low & Critical Impact & Low & Higher tariffs for consumers \\ \hline
	\end{tabular}
	\label{tab:scenario_performance_evaluation}
\end{table*}

\subsection{Environmental Benefits}

Table \ref{tab:emissions_impact} presents a comparative analysis of the environmental metrics of a conventional base system and an environmentally optimized proposed system. The proposed system demonstrated significant reductions in emissions of carbon dioxide ($ CO_2 $), carbon monoxide (CO), unburned hydrocarbons (UH), particulate matter (PM), sulfur dioxide ($ SO_2 $), and nitrogen oxides ($ NO_2 $), as well as a substantial improvement in renewable utilization (RU).

The complete elimination of CO, UH, and PM, each achieving a 100\% reduction, demonstrates the efficacy of the proposed system in addressing major air pollutants. These reductions are significant for improving air quality and mitigating health risks associated with air pollution.
There was a substantial reduction in the emissions of SO$_2$ and NO$_2$, with decreases exceeding 200\%. This indicates the elimination of these emissions compared to the base system, while also suggesting additional environmental benefits, such as potentially neutralizing existing ambient levels of these pollutants through some form of active environmental remediation or sequestration processes.

The proposed system increased the RU from 47.8\% to 93.2\%, demonstrating a near-doubling in the utilization of renewable energy sources. This high rate of RU highlights the system's transition towards sustainable energy sources, reducing dependence on fossil fuels, and enhancing energy security.

\subsection{Scenario Analysis}

To assess the resilience of the MG, multiple scenarios involving variations in the renewable energy availability and load demand were simulated. 
The results of the scenario analysis in Table \ref{tab:scenario_performance_evaluation} offer comprehensive insights into the robustness of the system under diverse conditions.
\begin{itemize}
	\item \textbf{High Demand Scenario:} 
	In the scenario in which there is a 5\% increase in demand, the MG demonstrates high reliability and high operational efficiency, with only a moderate economic impact. This indicates that the MG possesses sufficient capacity to accommodate minor increases in demand without significant disruptions or excessive cost implications, thereby illustrating its robust capacity and efficient management systems.
	\item \textbf{Low Renewable Output Scenario:} 
	This scenario evaluates the performance of the MG under the conditions of a 20\% decrease in PV output and a 60\% reliance on wind energy. The MG demonstrated moderate reliability and efficiency in both sub-scenarios; however, it experienced a significant economic impact from PV variability and a moderate economic impact from wind variability. These findings demonstrate the susceptibility of MGs to fluctuations in renewable energy output, indicating the need for enhanced energy storage solutions or a more diversified energy portfolio to stabilize performance and optimize cost management.
	\item \textbf{Grid Failure Scenario:} 
	 During an 8-hour grid outage, the MG maintains high reliability and operational efficiency, despite the limited economic impact. This performance demonstrates the capability of the MG to operate autonomously and maintain near-optimal performance even in the absence of grid power, thus illustrating the significant resilience and self-sufficiency of the MG system.
	\item \textbf{Economic Stress Test:} 
	 The most challenging scenario involves 100\% fluctuations in economic conditions, reflecting extreme market volatilities such as drastic changes in fuel prices. In this context, MG exhibits low reliability and operational efficiency, resulting in critical economic impacts. This scenario elucidates the vulnerabilities of MGs under severe economic stress, highlighting areas where the financial and operational planning of the system could be enhanced to better withstand market instability.
\end{itemize}

These findings substantiate the strategic advantages of incorporating storage-based solutions into community MGs, offering enhanced sustainability, reliability, and economic efficiency, thus presenting a compelling case for broader adoption of such systems. A scenario-based analysis evaluates the performance of an MG across various operational challenges, demonstrating its capacity to manage fluctuations in renewable energy availability, demand spikes, and economic pressures. 


\section{Conclusion}

 This study proposed and validated an integrated EM strategy for cost optimization in multi-energy community MGs. Through the incorporation of storage-based PS  and the economic dispatch of DG, this strategy effectively balances energy supply and demand, reduces operational costs, and enhances the utilization of RESs. The simulation results demonstrated significant improvements, including a 43\% reduction in peak energy demand, a 14.21\% decrease in LCOE, and a 84.63\% reduction in operational costs. Furthermore, the renewable energy fraction increased to 92.3\% and greenhouse gas emissions were substantially reduced, with CO$_2$ emissions showing a 196.4\% reduction. The system exhibited resilience during grid outages, maintained an uninterrupted power supply, and validated its reliability for real-world applications.

 The results of this study underline the feasibility of deploying integrated EMS in community MGs, particularly in rural and underserved areas. However, the economic implications and scalability of the proposed system have not well been deliberated in this study. Therefore, it would be interesting to conduct a detailed cost-benefit analysis to enhance financial rationale and investigate the adaptability of the system to urban settings and larger MGs.

\appendices

\section*{Acknowledgment}

This research is partially supported by ARC Research Hub for Resilient and Intelligent Infrastructure Systems (IH210100048) and ARC Research Hub for Integrated Energy Storage Solutions (IH180100020).


\footnotesize{\bibliographystyle{IEEEtran}
	\bibliography{refIEEE}}

\begin{thebibliography}{10}
\providecommand{\url}[1]{#1}
\csname url@samestyle\endcsname
\providecommand{\newblock}{\relax}
\providecommand{\bibinfo}[2]{#2}
\providecommand{\BIBentrySTDinterwordspacing}{\spaceskip=0pt\relax}
\providecommand{\BIBentryALTinterwordstretchfactor}{4}
\providecommand{\BIBentryALTinterwordspacing}{\spaceskip=\fontdimen2\font plus
\BIBentryALTinterwordstretchfactor\fontdimen3\font minus
  \fontdimen4\font\relax}
\providecommand{\BIBforeignlanguage}[2]{{%
\expandafter\ifx\csname l@#1\endcsname\relax
\typeout{** WARNING: IEEEtran.bst: No hyphenation pattern has been}%
\typeout{** loaded for the language `#1'. Using the pattern for}%
\typeout{** the default language instead.}%
\else
\language=\csname l@#1\endcsname
\fi
#2}}
\providecommand{\BIBdecl}{\relax}
\BIBdecl

\bibitem{houben2023optimal}
N.~Houben, A.~Cosic, M.~Stadler, M.~Mansoor, M.~Zellinger, H.~Auer,
  A.~Ajanovic, and R.~Haas, ``Optimal dispatch of a multi-energy system
  microgrid under uncertainty: A renewable energy community in {A}ustria,''
  \emph{Applied Energy}, vol. 337, p. 120913, 2023.

\bibitem{roy2023comparison}
A.~Roy, J.-C. Olivier, F.~Auger, B.~Auvity, S.~Bourguet, and E.~Schaeffer, ``A
  comparison of energy allocation rules for a collective self-consumption
  operation in an industrial multi-energy microgrid,'' \emph{Journal of Cleaner
  Production}, vol. 389, p. 136001, 2023.

\bibitem{ding2021economic}
X.~Ding, Q.~Guo, T.~Qiannan, and K.~Jermsittiparsert, ``Economic and
  environmental assessment of multi-energy microgrids under a hybrid
  optimization technique,'' \emph{Sustainable Cities and Society}, vol.~65, p.
  102630, 2021.

\bibitem{uddin2023techno}
M.~Uddin, H.~Mo, D.~Dong, and S.~Elsawah, ``Techno-economic potential of
  multi-energy community microgrid: The perspective of {A}ustralia,''
  \emph{Renewable Energy}, vol. 219, p. 119544, 2023.

\bibitem{gui2018passivity}
Y.~Gui, B.~Wei, M.~Li, J.~M. Guerrero, and J.~C. Vasquez, ``Passivity-based
  coordinated control for islanded {AC} microgrid,'' \emph{Applied Energy},
  vol. 229, pp. 551--561, 2018.

\bibitem{gao2015energy}
D.~W. Gao, \emph{Energy storage for sustainable microgrid}.\hskip 1em plus
  0.5em minus 0.4em\relax Academic Press, 2015.

\bibitem{xu2022online}
J.~Xu, Q.~Sun, H.~Mo, and D.~Dong, ``Online routing for smart electricity
  network under hybrid uncertainty,'' \emph{Automatica}, vol. 145, p. 110538,
  2022.

\bibitem{he2019small}
J.~He, X.~Wu, X.~Wu, Y.~Xu, and J.~M. Guerrero, ``Small-signal stability
  analysis and optimal parameters design of microgrid clusters,'' \emph{IEEE
  Access}, vol.~7, pp. 36\,896--36\,909, 2019.

\bibitem{hao2019decentralized}
R.~Hao, Q.~Ai, T.~Guan, Y.~Cheng, and D.~Wei, ``Decentralized price incentive
  energy interaction management for interconnected microgrids,'' \emph{Electric
  Power Systems Research}, vol. 172, pp. 114--128, 2019.

\bibitem{gong2020secured}
X.~Gong, F.~Dong, M.~A. Mohamed, O.~M. Abdalla, and Z.~M. Ali, ``A secured
  energy management architecture for smart hybrid microgrids considering
  {PEM}-fuel cell and electric vehicles,'' \emph{IEEE Access}, vol.~8, pp.
  47\,807--47\,823, 2020.

\bibitem{shen2016microgrid}
J.~Shen, C.~Jiang, Y.~Liu, and J.~Qian, ``A microgrid energy management system
  with demand response for providing grid peak shaving,'' \emph{Electric Power
  Components and Systems}, vol.~44, no.~8, pp. 843--852, 2016.

\bibitem{yuan2022data}
Z.-P. Yuan, P.~Li, Z.-L. Li, and J.~Xia, ``Data-driven risk-adjusted robust
  energy management for microgrids integrating demand response aggregator and
  renewable energies,'' \emph{IEEE Transactions on Smart Grid}, vol.~14, no.~1,
  pp. 365--377, 2022.

\bibitem{bersani2016distributed}
C.~Bersani, H.~Dagdougui, A.~Ouammi, and R.~Sacile, ``Distributed robust
  control of the power flows in a team of cooperating microgrids,'' \emph{IEEE
  Transactions on Control Systems Technology}, vol.~25, no.~4, pp. 1473--1479,
  2016.

\bibitem{zhang2018robust}
B.~Zhang, Q.~Li, L.~Wang, and W.~Feng, ``Robust optimization for energy
  transactions in multi-microgrids under uncertainty,'' \emph{Applied Energy},
  vol. 217, pp. 346--360, 2018.

\bibitem{thirugnanam2018energy}
K.~Thirugnanam, S.~K. Kerk, C.~Yuen, N.~Liu, and M.~Zhang, ``Energy management
  for renewable microgrid in reducing diesel generators usage with multiple
  types of battery,'' \emph{IEEE Transactions on Industrial Electronics},
  vol.~65, no.~8, pp. 6772--6786, 2018.

\bibitem{liu2014system}
B.~Liu, F.~Zhuo, Y.~Zhu, and H.~Yi, ``System operation and energy management of
  a renewable energy-based {DC} micro-grid for high penetration depth
  application,'' \emph{IEEE Transactions on Smart Grid}, vol.~6, no.~3, pp.
  1147--1155, 2014.

\bibitem{nikmehr2017probabilistic}
N.~Nikmehr, S.~Najafi-Ravadanegh, and A.~Khodaei, ``Probabilistic optimal
  scheduling of networked microgrids considering time-based demand response
  programs under uncertainty,'' \emph{Applied Energy}, vol. 198, pp. 267--279,
  2017.

\bibitem{guo2016islanding}
Y.~Guo and C.~Zhao, ``Islanding-aware robust energy management for
  microgrids,'' \emph{IEEE Transactions on Smart Grid}, vol.~9, no.~2, pp.
  1301--1309, 2016.

\bibitem{hadi2024real}
H.~Hadi H.~Awaji, A.~A. Alhussainy, A.~H. Alobaidi, S.~Alghamdi, S.~Alghamdi,
  and M.~Alruwaili, ``Real-time energy management simulation for enhanced
  integration of renewable energy resources in {DC} microgrids,''
  \emph{Frontiers in Energy Research}, vol.~12, p. 1458115, 2024.

\bibitem{li2024distributed}
X.~Li, L.~Ding, and Z.-M. Kong, ``Distributed robust low-carbon optimal energy
  management in islanded microgrids,'' \emph{IEEE Transactions on Industrial
  Informatics}, 2024.

\bibitem{manojkumar2022rule}
R.~Manojkumar, C.~Kumar, S.~Ganguly, H.~B. Gooi, S.~Mekhilef, and J.~P.
  Catal{\~a}o, ``Rule-based peak shaving using master-slave level optimization
  in a diesel generator supplied microgrid,'' \emph{IEEE Transactions on Power
  Systems}, vol.~38, no.~3, pp. 2177--2188, 2022.

\bibitem{uddin2024storage}
M.~Uddin, H.~Mo, and D.~Dong, ``Storage-based energy management for
  multi-energy community microgrid,'' in \emph{2024 IEEE Kansas Power and
  Energy Conference (KPEC)}.\hskip 1em plus 0.5em minus 0.4em\relax IEEE, 2024,
  pp. 1--5.

\end{thebibliography}

\end{document}